\documentclass[useAMS]{mn2e}
\title{Near-Infrared Studies of   V1280 Sco (Nova Scorpii 2007)}

\author[R.K. Das, D.P.K. Banerjee, N.M. Ashok, $\&$ O. Chesneau]{R.K. Das,$^{1}$
\thanks{E-mail: rkdas@prl.res.in (RKD); orion@prl.res.in (DPKB); ashok@prl.res.in (NMA);
olivier.chesneau@ob-azur.fr (OC)}
D.P.K. Banerjee$^{1}$, N.M. Ashok$^{1}$ $\&$ O. Chesneau$^{2}$\\
$^{1}$Astronomy $\&$ Astrophysics Division, Physical Research Laboratory, Navrangpura, Ahmedabad 380009, India\\
$^{2}$UMR 6525 H. Fizeau, Univ. Nice Sophia Antipolis, CNRS, Observatoire de
la C\^{o}te d'Azur,
Av. Copernic, F-06130 Grasse, France}

\usepackage{graphicx}
\begin{document}

\pagerange{\pageref{firstpage}--\pageref{lastpage}} \pubyear{2008}

\maketitle

\label{firstpage}

\begin{abstract}

We present spectroscopic and photometric results of Nova V1280 Sco which was discovered
in outburst in early 2007 February. The large number of spectra obtained of the object
leads to one of the most extensive, near-infrared spectral studies of a classical
nova. The spectra evolve from a P-Cygni phase to an emission-line
phase and at a later stage is dominated by emission from the dust that formed in this
nova. A detailed model is computed to identify and study characteristics of the spectral
lines. Inferences from the model address the vexing question of which
novae have the ability to form dust. It is demonstrated, and strikingly corroborated
with  observations,  that the presence of lines in the early spectra of low-ionization species like Na
and Mg - indicative of low temperature conditions - appear to be reliable indicators
that dust will form in the ejecta. It is theoretically expected that mass loss  during
a nova outburst
is a sustained process. Spectroscopic evidence for such a sustained mass loss, obtained by tracing the evolution of a P-Cygni feature in the Brackett $\gamma$ line, is presented here allowing a
lower limit of 25-27 days to be set for the mass-loss duration. Photometric data recording the
nova's extended 12 day climb to peak brightness after discovery is used to establish
an early fireball expansion
and also show that the ejection began well before maximum brightness. The JHK light curves
indicate the nova  had a fairly strong second outburst $\sim$ 100 days after the
first.

\end{abstract}


\section{Introduction}
Nova V1280 Sco was discovered on 2007 February 4.8  independently by Sakurai
and Nakamura  (2007), within  a short time of each other, with a reported visual magnitude  in the range  9.4 - 9.9.  The nova brightened quickly
to its maximum in visual light on Feb. 16.19 ($m_{vmax}$= 3.79, Munari et al.,
2007) to become one of the brightest novae in recent times ( Schmeer 2007); V1500 Cyg
in 1975 had a $m_{vmax}$=2.0 and nova V382 Vel 1999 had $m_{vmax}$ =2.5.
V1280 Sco has been observed in different wavelength regimes since its discovery.
Early stage pre-maxima spectra was described by Munari et al. (2007) and
Naito $\&$ Narusawa (2007) in the visible region and by Rudy et
al. (2007a,b) in the infrared. The early spectrum was dominated by
absorption lines of hydrogen, neutral nitrogen and carbon, and displayed deep P-Cygni
profiles. After passing the maxima the spectrum turned to that typically shown by a
classical nova; the lines started appearing in emission which included strong
emission lines of Ca II H and K, the hydrogen Balmer series, the Na I D doublet,
and Fe II 42, 49 and 74 in visible region (Buil, 2007; Munari et al., 2007). Early
near-IR spectra by  Das et al. (2007a) reported the presence of HI
Brackett and Paschen lines and several strong C I lines. No X-ray detection was
reported from the source from RXTE and SWIFT observations respectively (Swank et al. 2007; Osborne et al., 2007) .

The nova has an interesting lightcurve with  the possibility of more than one
outburst. After passing the maxima, a  smooth and slow decline followed. This was
interrupted, about 24 days after discovery, by the formation of dust
which was evidenced from the sharp decline in visible light curve and a rise
in the infrared continuum (Das et al., 2007b). The dust shell has been
directly detected by interferometric techniques  and its subsequent spatial
expansion has also been tracked (Chesneau et al. 2008). Apart from providing
the first direct detection of a dust shell around a classical nova, the Chesneau
et al. (2008) study emphasises the  important, emerging role of interferometry in
the study of dust formation and dust properties in novae. We present here spectroscopic and photometric
results based on  observations between 14 and 125 days after the discovery. A
large number of spectra were recorded, sampling the nova's evolution
at regular intervals, thereby leading to one of the most comprehensive studies
of a  classical nova in the near-infrared.

\section{Observations}
Observations of V1280 Sco in Near-IR $JHK$ were obtained at the Mt. Abu 1.2m
telescope. Near-IR $JHK$ spectra presented here were obtained at  similar dispersions of $\sim$ 9.75 {\AA}/pixel in each of the $J,H,K$ bands using the Near Infrared
Imager/Spectrometer with a 256$\times$256 HgCdTe NICMOS3 array. Generally, a set
of at least two spectra were taken with the object dithered to two positions
along the slit.  The spectra were extracted using IRAF and wavelength calibration
was done using a combination of  OH sky lines and telluric lines
that register with the stellar spectra. Following the standard procedure, the
nova spectra were then ratioed with the  spectra of a comparison star
(SAO 184301; spectral type A0V) from whose spectra the Hydrogen Paschen and Brackett
absorption
lines had been extrapolated out. The lack of a suitably bright standard star
close to the nova, thereby leading to our choice SAO 184301, led to some
difference in airmass  between the nova and standard star observations.
Thus, the ratioing process, while removing telluric features sufficiently well, does
leave some residuals. This applies to  regions where telluric absorption is strong (specifically
the 1.12 ${\rm{\mu}}$m region in the $J$ band and the 2 to 2.05 ${\rm{\mu}}$m region in the
$K$ band affected by atmospheric oxygen and cabon-dioxide respectively; we have thus excluded the latter $K$ band region from our spectra). Photometry in
the $JHK$ bands was done  in photometric sky conditions using the imaging mode
of the NICMOS3 array. Several frames,  in 5 dithered positions offset typically by
20 arcsec, were obtained of both the nova and a selected standard star in each of the
$J,H,K$ filters. Near-IR $JHK$ magnitudes were then derived using IRAF tasks and
following the regular procedure followed by us for photometric reduction (e.g. Banerjee $\&$ Ashok, 2002). The log of the photometric observations and the derived $JHK$
magnitudes, with typical errors in the range 0.01 to 0.03 magnitudes, are given in Table 1.

\begin{table}
\centering
\caption{A log of the photometric observations of V1280 Sco. The date of outburst
is taken to be 2007 Feb 4.854 UT}

\begin{tabular}{lccccc}
\hline
Date & Days         & &             & Magnitudes&          \\
2007 & since        & &             &           &          \\
(UT) & Outburst     & & \emph{J}    & \emph{H}  & \emph{K} \\
\hline
\hline
Feb. 19.972  & 15.118 & & 3.33    & 3.05     & 2.78   \\
Feb. 25.962  & 21.108 & & 3.58    & 3.09     & 2.77  \\
Feb. 27.991  & 23.137 & & 4.00    & 3.63     & 3.79  \\
Mar. 01.992  & 25.138 & & 4.71    & 4.27     & 4.21  \\
Mar. 03.996  & 27.142 & & 5.42    & 4.43     & 3.55  \\
Mar. 05.991  & 29.137 & & 6.23    & 4.88     & 3.62  \\
Mar. 07.995  & 31.141 & & 6.68    & 5.09     & 3.78  \\
Mar. 10.986  & 34.132 & & 6.82    & 5.15     & 3.93  \\
Mar. 13.979  & 37.125 & & 7.21    & 5.45     & 4.03  \\
Mar. 27.950  & 51.096 & & 7.62    & 5.51     & 3.88  \\
Mar. 28.971  & 52.117 & & 7.72    & 5.61     & 3.83  \\
Mar. 31.981  & 55.128 & & 8.25    & 5.98     & 4.13  \\
Apr. 01.972  & 56.118 & & 8.29    & 6.04     & 4.19  \\
Apr. 02.981  & 57.127 & & 8.25    & 5.97     & 4.01  \\
Apr. 03.988  & 58.134 & & 8.32    & 6.04     & 3.90  \\
Apr. 05.975  & 60.121 & & 8.27    & 6.08     & 4.20  \\
Apr. 08.982  & 63.128 & & 8.27    & 6.05     & 4.05  \\
Apr. 15.963  & 70.109 & & 8.10    & 5.90     & 3.91  \\
Apr. 18.953  & 73.099 & & 7.89    & 5.78     & 3.93  \\
Apr. 26.871  & 81.017 & & 8.84    & 6.58     & 4.52  \\
Apr. 30.890  & 85.04 & & 8.85     & 6.70     & 4.62  \\
May  02.920  & 87.066 & & 8.69    & 6.38     & 4.36  \\
May  03.922  & 88.068 & & 9.10    & 6.71     & 4.49  \\
May  04.899  & 89.045 & & 9.33    & 6.95     & 4.81  \\
May  05.920  & 90.066 & & 9.06    & 6.79     & 4.63  \\
May  06.898  & 91.044 & & 8.64    & 6.38     & 4.26  \\
May  08.910  & 93.056 & & 9.13    & 6.77     & 4.62  \\
May  15.798  & 99.944 & & 8.81    & 6.41     & 4.11  \\
May  22.800  & 106.946& & 7.26    & 4.94     & 3.06  \\
May  30.811  & 114.957& & 7.36    & 5.16     & 3.25  \\
Jun. 02.809  & 117.955& & 7.73    & 5.47     & 3.23  \\
Jun. 05.791  & 120.937& & 8.08    & 5.73     & 3.67  \\
Jun. 07.772  & 122.918& & 8.23    & 5.78     & 3.58  \\
Jun. 10.773  & 125.919& & 8.15    & 5.77     & 3.81  \\
\hline
\end{tabular}
\end{table}

\section{Results}
\subsection{Optical lightcurve} Before presenting the results proper, we estimate
some of the  useful parameters for V1280 Sco. The optical lightcurve is presented in
Figure 1. As  seen, V1280 Sco is one of those few  novae in which the
light-curve data in the pre-maximum stage is well documented -  two other novae,
which come readily to mind, in which the pre-maximum ascent is similarly well
studied and used to determine the onset of eruption are V1500 Cyg and PW Vul (Gehrz 1988 and references therein). After the maximum, the
light curve declined steadily but was interrupted by dust formation
around 24 days after discovery. We  discuss in detail subsequently  the dust
formation phase and other specific  phases in the light-curve evolution. For the
present we wish to determine $t$${_{\rm 2}}$, the time for a decline
of two magnitudes in the visual band, and note  that the light curve indicates that
dust formation had begun even before $t$${_{\rm 2}}$ was reached. Hence to estimate $t$${_{\rm 2}}$, we have
extrapolated the light-curve linearly in the post-maximum decline stage, as would have
been the case had dust not formed at all, and thereby estimate $t$${_{\rm 2}}$ to be
approximately 21 days though slightly smaller values may also be supported. Using the MMRD relation of Della Valle and Livio (1995) we
determine the absolute magnitude of the nova to be M${_{\rm v}}$ = 7.88 and obtain a
 distance estimate  to
the object $d$ = 1.25 kpc for an assumed value of A${_{\rm v}}$ = 1.2 as inferred from
the extinction in the direction of V1280 Sco from the work of Marshall et al. (2006). It should be noted that the estimated value of $d$ could be subject to significant uncertainty.
This arises from errors in estimating  $t$${_{\rm 2}}$, A${_{\rm v}}$ and also from the intrinsic errors associated with the  MMRD relations. \\

Henden $\&$ Munari (2007) have made  post-ouburst astrometric and  photometric measurements  of the stars in the field around V1280 Sco and generated a photometric sequence to calibrate archival plates. Using SuperCosmos plates, a comparison was made  between  the magnitudes of stars therein with those of Henden $\&$ Munari (2007). A good consistency is found which allows us to conclude from the SuperCosmos data that no star
is seen at the nova's position down to  $B$ and $R$ magnitudes of 20.3 and 19.3 respectively. A lower limit on the outburst amplitude $A$ of the nova in the visual
region can therefore be estimated to be in the range of $\sim$ 15.5 to 16 magnitudes.
Apart from the  amplitude of the outburst being rather large,  the compiled data of $A$
versus $t$${_{\rm 2}}$ for different novae (Warner, 2008) show that V1280 Sco is very much an outlier. This, together with its long rise to maximum, early dust formation and
the the presence of a late-time, secondary outburst makes the object somewhat unusual.

\begin{table}
\caption{A log of the spectroscopic observations of V1280 Sco. The date of outburst
is taken to be 2007 Feb 4.854 UT}

\begin{tabular}{lcccccc}
\hline\
Date & Days        & &         & Integration time &   \\
2007 & since        & &        & (sec)            &    \\
(UT) & Outburst  & & \emph{J} & \emph{H}         & \emph{K}    \\
\hline
\hline
Feb. 18.981  & 14.127   & & 10      & 10               & 10 \\
Feb. 20.006  & 15.152   & & 10      & 10               & 10 \\
Feb. 24.990  & 20.136   & & 10      & 10               & 10 \\
Feb. 25.997  & 21.143   & & 10      & 10               & 10 \\
Feb. 26.968  & 22.114   & & 10      & 8                & 10 \\
Feb. 27.972  & 23.118   & & 10      & 8                & 10 \\
Mar. 01.964  & 25.110   & & 10      & 15               & 15 \\
Mar. 03.961  & 27.107   & & 20      & 15               & 20 \\
Mar. 04.961  & 28.107   & & 20      & 20               & 20 \\
Mar. 05.965  & 29.111   & & 45      & 20               & 20 \\
Mar. 06.965  & 30.111   & & 40      & 30               & 30 \\
Mar. 07.951  & 31.097   & & 60      & 20               & 20 \\
Mar. 10.963  & 34.109   & & 90      & 45               & 30 \\
Mar. 13.950  & 37.096   & & 100     & 40               & 30 \\
Mar. 28.928  & 52.074   & & 150     & 60               & 30 \\
Mar. 31.920  & 55.066   & & 300     & 45               & 30 \\
Apr. 02.902  & 57.048   & & 300     & 40               & 30 \\
Apr. 05.905  & 60.051   & & 350     & 60               & 40 \\
Apr. 18.891  & 73.037   & & 500     & 60               & 30 \\
Apr. 27.870  & 82.016   & & 500     & 60               & 45 \\
May. 04.954  & 89.100   & & 450     & 60               & 45 \\
May. 06.833  & 90.979   & & 500     & 45               & 30 \\
May. 22.865  & 107.011   & & 300     & 40               & 20 \\
May. 30.854  & 114.999   & & 350     & 40               & 30 \\
Jun. 05.758  & 120.904    & & 400     & 90               & 30 \\
Jun. 08.836  & 123.982    & & 350     & 60               & 40 \\

\hline
\end{tabular}
\end{table}

\begin{figure}
\centering
\includegraphics[bb= 1 14 506 353, width=3.5in,height=3.0in,clip]{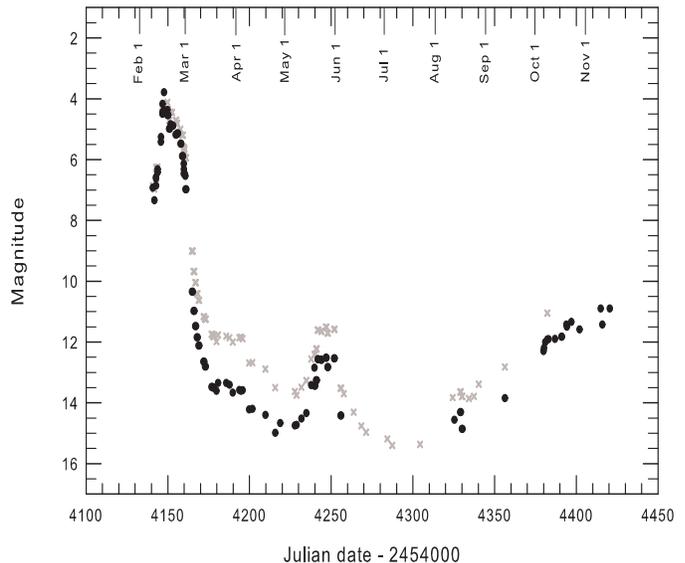}
\caption[]{ The $V$ and $R$${_{\rm c}}$ light curves (black circles and gray crosses respectively) of V1280 Sco from data obtained from IAU circulars,  VSNET alerts, VSOLJ and AFOEV databases ((Variable Stars Network, Japan;
Variable Star Observers League in Japan; Association Francaise des Observateurs
d'Etoiles Variables, France)}
\label{fig1}
\end{figure}


\subsection{General characteristics of the $JHK$ spectra} The log of observations for
the spectroscopy is presented in Table 2. Instead of presenting all the spectra,
 we have rather chosen representative spectra  sampling the observations at regular intervals which effectively convey the trend/behavior  of the spectral evolution.
These
$JHK$ spectra are presented in Figures 2, 3, and 4 respectively.
The earliest spectra of 18 Feb show the lines to display P-Cygni structures with
the absorption components
considerably stronger than the emission components which are just beginning to
appear. This is consistent with the near-IR spectroscopy of Rudy et al. (2007b)
between 14 and 16 Feb. wherein they report that the lines of hydrogen were
in absorption;  emission lines  almost entirely absent and Pa $\beta$ and
Br $\alpha$ were just beginning to exhibit hints of emission and
P-Cygni behavior.  By 19 Feb, the emission
components of the P-Cygni profiles become stronger than the absorption components
and by 24 February the
spectra become of pure emission type typical of a classical nova soon after maximum light. The emission lines remain at significant strength thereafter
till the first week of March after which  they lose contrast
against the rising continuum from dust emission. The formation of dust in
the ejecta at around the beginning of 2007 March can also be inferred - apart from
it being seen as a drop in the visible light curve - from the change seen in the slope of the near-IR continuum. In each of the $JHK$ bands a rise in the continuum
towards longer wavelengths is clearly seen. The development of this infrared
excess, attributable to dust, is seen till the end of our observations   indicating that the freshly-formed dust persists till then or is augmented by
further episodes of dust formation. A detailed investigation of these aspects
and analysis of the dust shell properties and kinematics is given in Chesneau et al. (2008).

\subsection{Line identification and detailed study of spectral features}

To facilitate a better understanding of the lines that contribute to a
nova's spectrum,  a  model is developed based on local thermal equilibrium (LTE) considerations. It is likely that the model has limitations, based as it is
on LTE assumptions which may not strictly prevail in an nova environment. Yet, we show that the model-generated spectra, greatly aids
in a  secure identification of the lines observed and also gives additional
valuable insights. The synthetic spectrum is computed  along similar lines as for nova V445 Pup in Ashok $\&$ Banerjee (2003) but is extended further
here. The model spectra are generated by considering only those elements whose
lines can be expected at discernible strength. Since   nucleosynthesis calculations of elemental abundances in novae  (Starrfield et al. 1997; Jose $\&$ Hernanz, 1998) show that H, He, C, O, N, Ne, Mg, Na, Al, Si, P, S  are the elements with significant yields in novae ejecta, only these elements have been considered.
For a particular element, we have first used the Saha ionization equation to calculate the fractional percentage of the species in different ionization stages (neutral plus higher ionized states). Subsequently the Boltzman equation is applied to calculate the population of the upper level from which a transition of  interest arises.  The transitions of interest are essentially
all the stronger
lines of the above elements in the region of interest (1-2.5 ${\rm{\mu}}$m) region which was
compiled from the Kurucz atomic line list{\footnote {http://cfa-www.harvard.edu/amp/ampdata/}} and National Institute of Standards
and Technology (NIST){\footnote {http://physics.nist.gov/PhysRefData/ASD}}  line list database.
From the database, for each line the  statistical
weight  of the upper level of the transition, the energy difference
between the upper energy level  and the ground state; and  the
transition probability is also noted.  Given these parameters,
and the partition functions (adopted from Allen 1976; Aller 1963),  the
population of the upper level for a particular transition - under
LTE conditions - can then be derived  from the Boltzman equation
and the line strength can subsequently be calculated knowing the
transition probability. For simplicity we assume that the shape of each line can
be reasonably represented by a Gaussian whose strength is known
from the computed line strength and whose width can be adjusted to
match that of the observed profiles. The co-addition of all such
Gaussians - corresponding to all the lines - yields a model spectrum.
In computing a model spectrum, values of certain parameters need to be
assumed viz. the electron density $N$${_{\rm e}}$ (needed in the ionization equation); the gas temperature $T$ (needed in both the Saha and Boltzmann equations) and the
assumed abundances of the elements. We have considered typical values
found in nova ejecta in the early stages viz. $N$${_{\rm e}}$ in the range
10${^{\rm 9}}$ to 10${^{\rm 11}}$ cm${^{\rm -3}}$, $T$ $\sim$ 4000 - 10000K and abundances  in
line with those of CO noave given in Starrfield et al. (1997) and Jose $\&$
Hernanz (1998).\\

The computed spectrum is presented,  overlaid with a representative observed spectrum for comparison, in Figure 5 with the lines identified. The list of identified lines is given in Tables 3 and 4. While most of the lines observed are known and identified from previous studies, there are a few subtle aspects influencing specific lines  which are likely to have gone  unnoticed or whose significance could be under emphasized. A better understanding of these nuances emerge from our analysis. One such aspect is  determining the relative contribution of diffrent species to a particular spectral feature/line. Since in our model, trial spectra can be computed for one element at a time, the position and strength of all the lines of this element can be determined. Thus it becomes clear to assess whether  a particular line, at a particular wavelength, has one or more species is contributing to it. On the basis of such an analysis we arrive at the following conclusions:

1) In the $J$ band there is a Mg I line at 1.1828 ${\rm{\mu}}$m in the wing of the strong Carbon lines at
$\sim$ 1.1750 ${\rm{\mu}}$m.
This Mg I line, in case the nova emission lines are broad in general, may blend  with the C I feature at 1.1750 ${\rm{\mu}}$m giving the latter a broad redward wing. Alternatively
it may be  seen as a distinct spectral feature as  seen in V1419 Aql (Lynch et al. 1995)
or in  our observations on later days (e.g. the spectra of 3 and 5 March). Establishing
the identity of Mg lines unambiguously, and also that of Na lines, is important because we show subsequently that they are potential predictors of dust formation.

2) The region between 1.2 to 1.275 ${\rm{\mu}}$m is a complex blend of a very large number of  lines - principally those of NI and C I. This complex  is routinely seen - always  at  low strength - in the early spectra of several nova including
V2274 Cyg and V1419 Aql. Since our resolution in this region is higher than that of the
V2274 Cyg and V1419 Aql spectra, we have tried to identify  individual lines in the complex. Since most of the individual spectral features in this region are weak it is necessary to ensure their reliability. Thus, the data of 5 Feb was examined - since four, high S/N spectra are available for this date - and it was ensured that
individual features repeat in all the four spectra. Subsequent line identification that followed, by comparing with the synthetic spectrum, is largely satisfactory but not completely so. The majority of the observed features are reproduced at the correct wavelengths in the synthetic spectrum except for the observed  1.2074,1.2095 ${\rm{\mu}}$m C I blend and the unidentified line at 1.2140 ${\rm{\mu}}$m which are not. The NI 1.2461, 1.2569 ${\rm{\mu}}$m feature is one of the stronger features in the complex and while it has been attributed solely to NI by Rudy et al. (2003), it is likely that there is some contamination of this line with O I 1.2464 ${\rm{\mu}}$m. In fact, the excess strength shown by this line in the synthetic spectrum is due to the contribution of the O I line in addition to the NI line. The contribution of O I however has made the line look too strong.

3) The blended feature comprising of the O I 1.1287 ${\rm{\mu}}$m and C I 1.1330 ${\rm{\mu}}$m lines shows considerable evolution
with time. In the initial phase, for  spectra between  24 Feb and $\sim$ 5-7 March, the O I and C I lines contribute to the blend in comparable amounts. But with time the O I
line begins to dominate  possibly due to a increase of the Ly $\beta$ flux with time as the central remnant becomes hotter. That Ly $\beta$ fluorescence is certainly influencing the 1.1287 ${\rm{\mu}}$m line can be inferred from its large relative strength compared  to the continuum excited 1.3164 O I line. The 1.1287 ${\rm{\mu}}$m O I - C I 1.1330 ${\rm{\mu}}$m blended feature shows a broad red wing at 1.14 ${\rm{\mu}}$m. There appears to be strong case to believe, from the results of our synthetic spectrum, that Na I lines at 1.1381 and 1.1404 ${\rm{\mu}}$m are responsible for this red wing. There are two Carbon lines in this region which could contaminate the Na I lines but the effect of these C I lines is found to be very marginal from our model spectrum. It is also noted that the same Na I features are seen in the spectrum of   V2274 Cen and identified by Rudy et al. (2003) as potentially arising from Na I.

4) In the $H$ band, the recombination lines of the Brackett series are most prominent and
readily identifiable. But the presence of a C I line at 1.6 ${\rm{\mu}}$m, which
could  be mistaken as just another member of the Brackett series, should not
be missed. If a recombination analysis of the Hydrogen lines is to be done, caution
is needed in estimating line strengths of Br 11 which can be severely blended with
the strong C I feature at 1.6890 ${\rm{\mu}}$m and Br10 at  1.7362 ${\rm{\mu}}$m which is again blended
with C I lines. Other Br lines, whose strengths are affected but to much lesser extent,
are Br14 at  1.5881 ${\rm{\mu}}$m (blended with C I 1.5853 ${\rm{\mu}}$m and O I 1.5888 ${\rm{\mu}}$m ) thereby making it appear
artificially stronger than Br12 contrary to what is expected in a Case B scenario;
Br12 at 1.6407 ${\rm{\mu}}$m is contaminated with several weak C I lines between
1.6335 ${\rm{\mu}}$m and 1.6505 ${\rm{\mu}}$m thereby
giving Br12 a broad appearance on both  wings. A specifically interesting feature
at 1.579 ${\rm{\mu}}$m  on the redward flank of Br15 (1.5701 ${\rm{\mu}}$m  is due to Mg I;
magnesium and sodium lines are discussed subsequently.\\

\begin{figure*}
\centering
\includegraphics[bb=0 0 695 772,width=7.0in,height=4.0in]{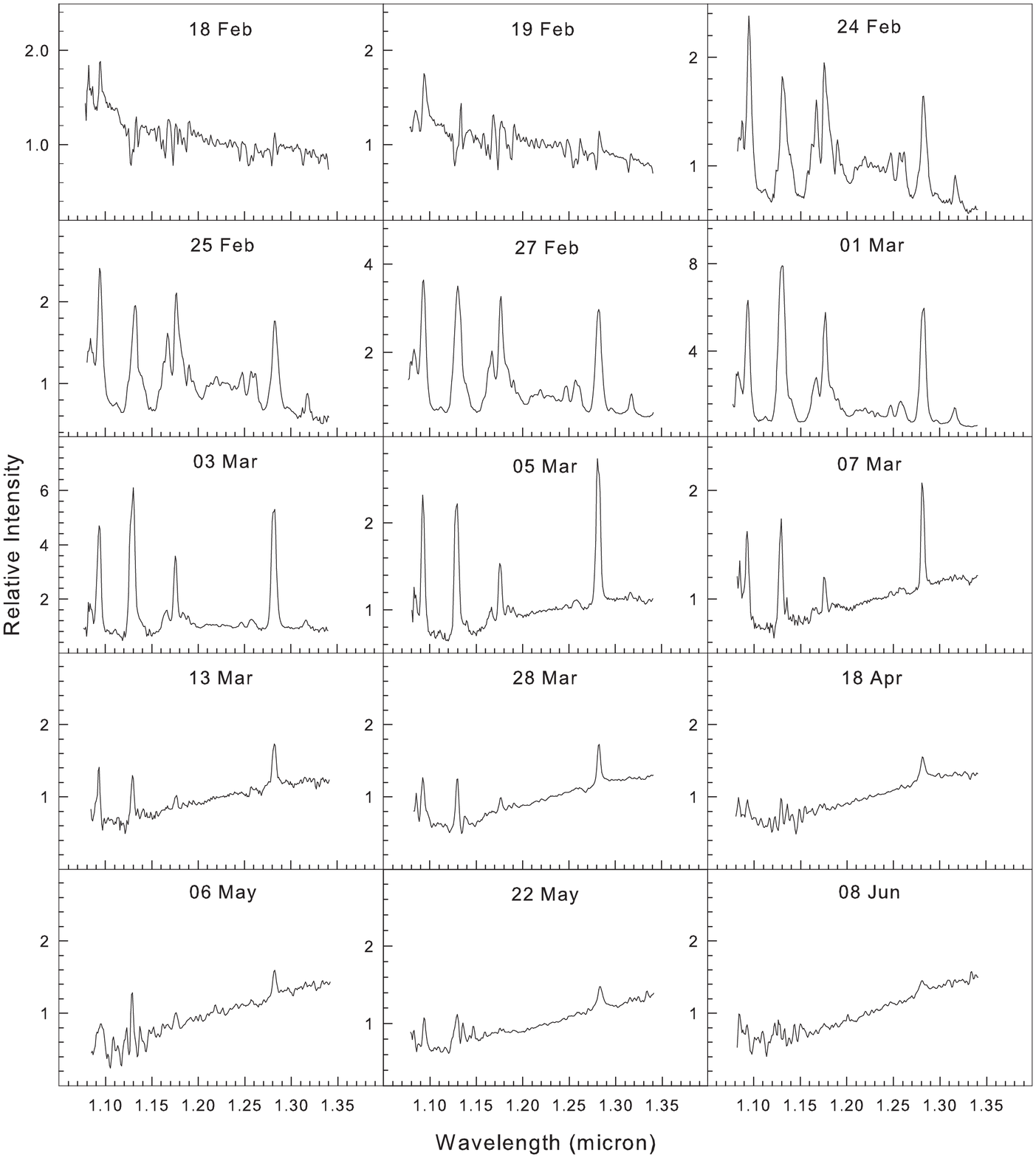}
\caption[]{ The $J$ band spectra of V1280 Sco on different days with the flux normalized to unity at 1.25 ${\rm{\mu}}$m.
}
\label{fig2}
\end{figure*}

\begin{figure*}
\includegraphics[bb=0 0 701 771,width=7.0in,height=4.0in]{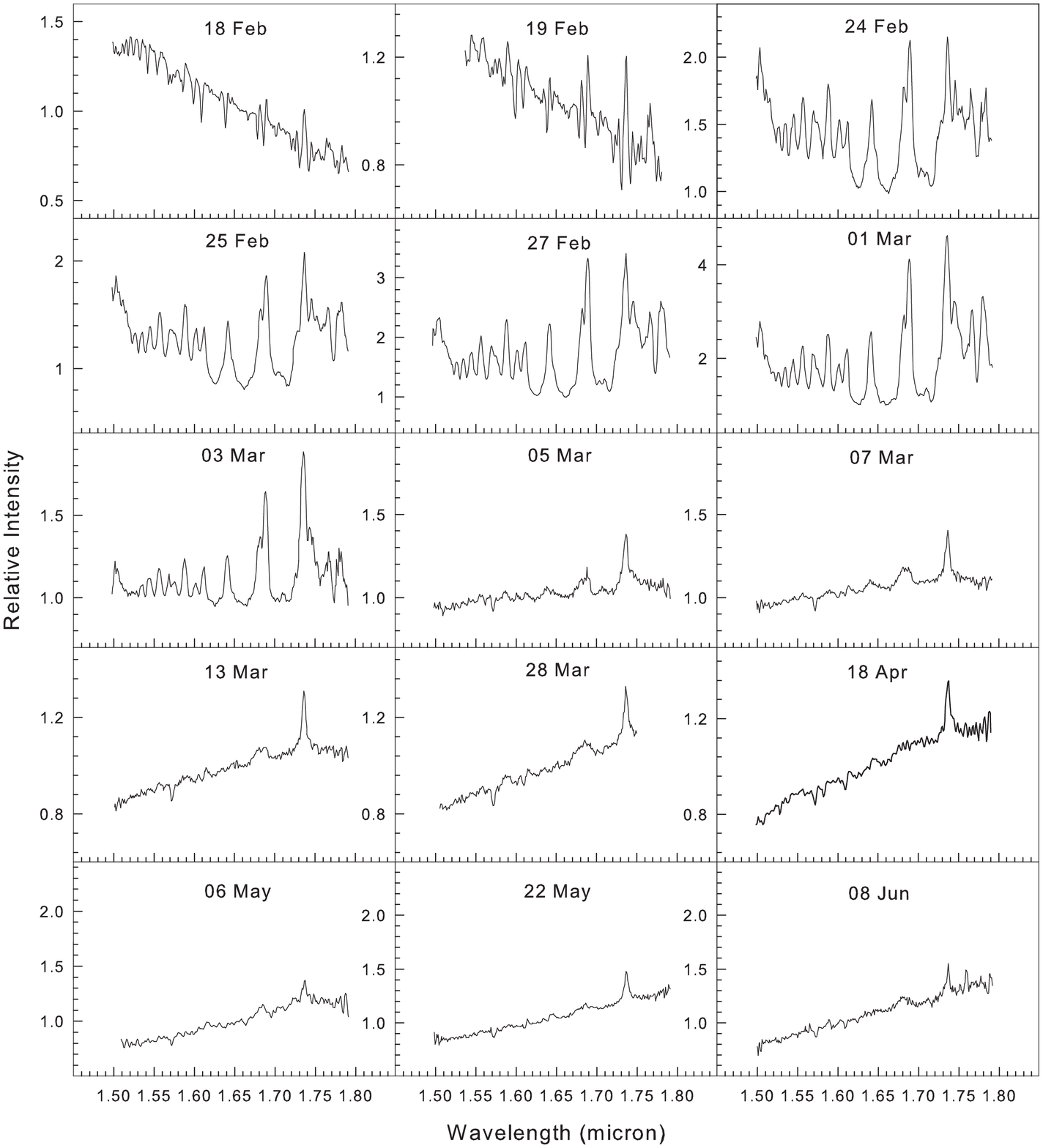}
\caption[]{ The $H$ band spectra of V1280 Sco on different days with the flux normalized to unity at 1.65 ${\rm{\mu}}$m.}
\label{fig2}
\end{figure*}

\begin{figure*}
*\centering
\includegraphics[bb=0 0 703 774,width=7.0in,height=3.8in,clip]{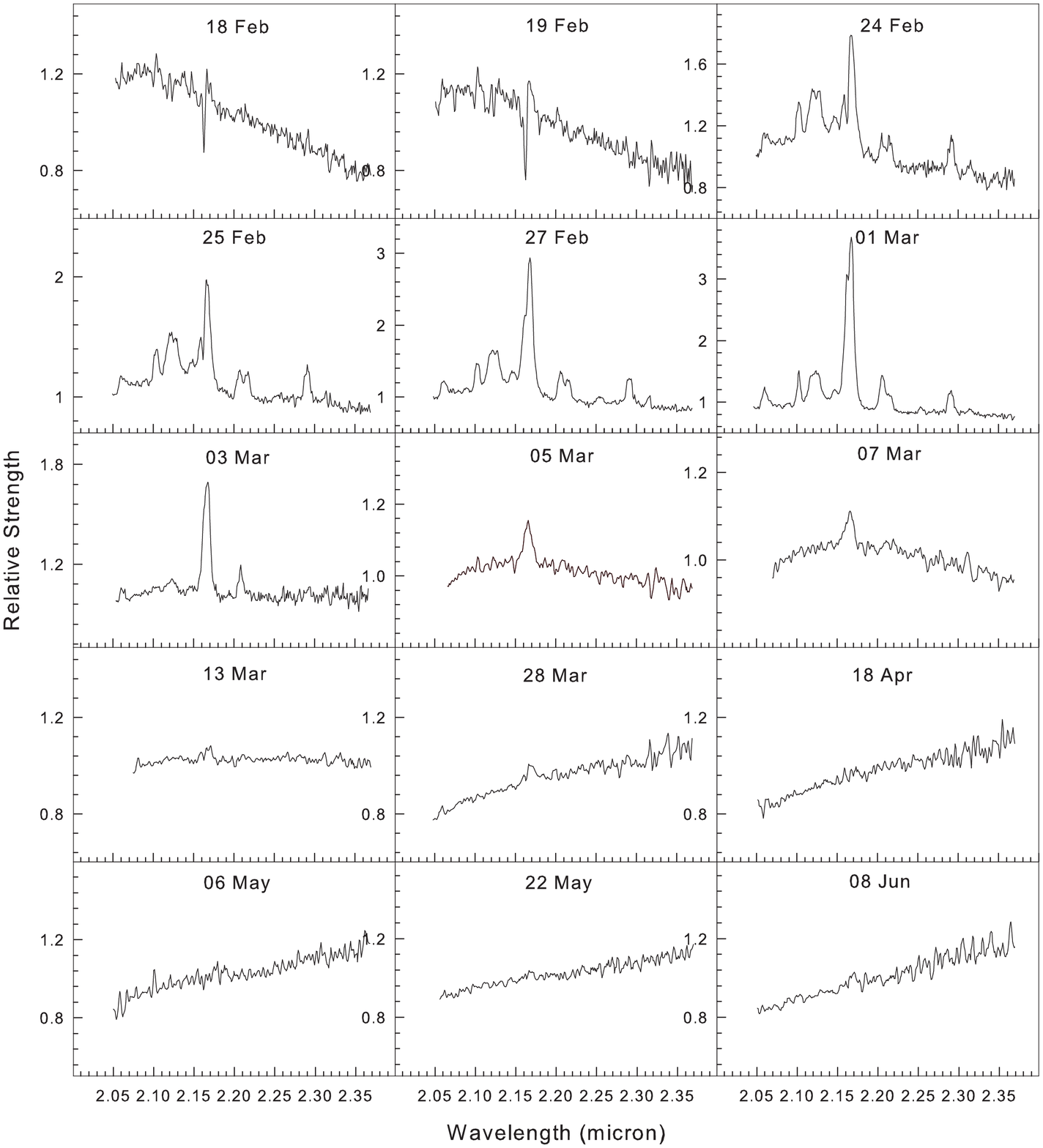}
\caption[]{ The $K$ band spectra of V1280 Sco on different days with the flux normalized to unity at 2.2 ${\rm{\mu}}$m.
}
\label{fig2}
\end{figure*}


Before proceeding with other  conclusions regarding line identification,
it is relevant to point out at this stage that the early spectrum of V1280 Sco
is very similar to that of the dust forming novae  V2274 Cyg and V1419 Aql
(Rudy et al. 2003, Lynch et al. 1995). A comparison of the 0.75-1.35  ${\rm{\mu}}$m early
spectra of these last two novae is given in Figure 8 of Rudy et al. (2003) to show
the striking similarity in spectral features between them   and
examination of the $J$ band spectra shows  that there is a one-to-one replication of spectral
features in the case of  V1280 Sco also. All three novae
show a spectrum rich in H,C, O and N. Rudy et al. (2003) discuss in detail why it
is not always easy to distinguish between CO and ONeMg novae based on near-infrared
spectroscopy alone. But  they show that by considering several observational features
together viz. the rate of decline, the expansion velocity, the formation
of dust and the presence of C I lines that V2274 Cyg is strongly indicated to have
a CO white dwarf (WD). By virtue of the same arguments, it would appear that V1280 Sco is
also a CO nova. This is further supported by the  classification of
V1280 Sco as a FeII nova based on its early optical spectrum (Munari et al. 2007) -
Fe II novae are associated  with explosions on CO white dwarfs (Williams 1992).
It is intended to present results on three other novae which we have studied recently viz.
V476 Scuti, V2615 Oph and V2575 Oph (Das et al. 2008; paper under preparation).
All three
show rather similar spectra as V1280 Sco with prominent carbon lines and furthermore
they all went on to form dust - a propensity shown preferentially by CO novae (Starrfield et al. 1997;  Gehrz  1988).  It is therefore
tempting to think that a tentative hypothesis is emerging, subject
to proper validation, that the characteristic near-IR spectrum shown
by these nova is a hallmark of the CO class of novae. On a more
definitive note, we hope to convincingly show that there are certain
features in the spectrum of V1280 Sco that appear to be reliable, generic predictors
whether dust will form in novae ejecta or not. A return is now made
to a few other  traits in the spectrum in Figure 5 that deserve elaboration.\\

5) Apart from the 1.1828 ${\rm{\mu}}$m  line, the other Mg I lines that we  detect are
those at 1.5040,
1.5749 and 1.7109 ${\rm{\mu}}$m . The 1.5040 ${\rm{\mu}}$m and 1.7109 ${\rm{\mu}}$m are also suggested by Rudy et al (2003)
to be due to Mg I - we are  convinced that this association is correct even if the
1.5040   ${\rm{\mu}}$m line is at the edge of our observed spectral window. The
 feature at 1.5749 ${\rm{\mu}}$m, blending with
the wings of Br 15 is certainly Mg I and may escape detection at lower resolutions where it
could blend with lines of the Brackett series. The 1.5749 ${\rm{\mu}}$m line can become quite strong, stronger than the adjacent Br lines as in the case of V2615 Oph (Das et al. 2008) and a
correlated increase in the strength between this line and other Mg I lines is also
clearly seen in V2615 Oph.
The identification of the Mg I lines is therefore felt to be  secure. Regarding lines
arising from Na, apart from those at $\sim$ 1.4 ${\rm{\mu}}$m , the  other lines that are
definitively identified are those at 2.2056 and 2.2084 ${\rm{\mu}}$m . These lines are known
to occur in novae e.g in V705 Cas (Evans et al. 1996). However, in the synthetic spectrum,
we have trouble reproducing  the 2.1045 ${\rm{\mu}}$m feature - again  associated
with Na I (e.g. Evans et al. 1996) - at  its observable strength. The identification
is therefore uncertain.

\begin{figure}
\centering
\includegraphics[bb=0 0 259 556,width=3.5in,height=8.0in,clip]{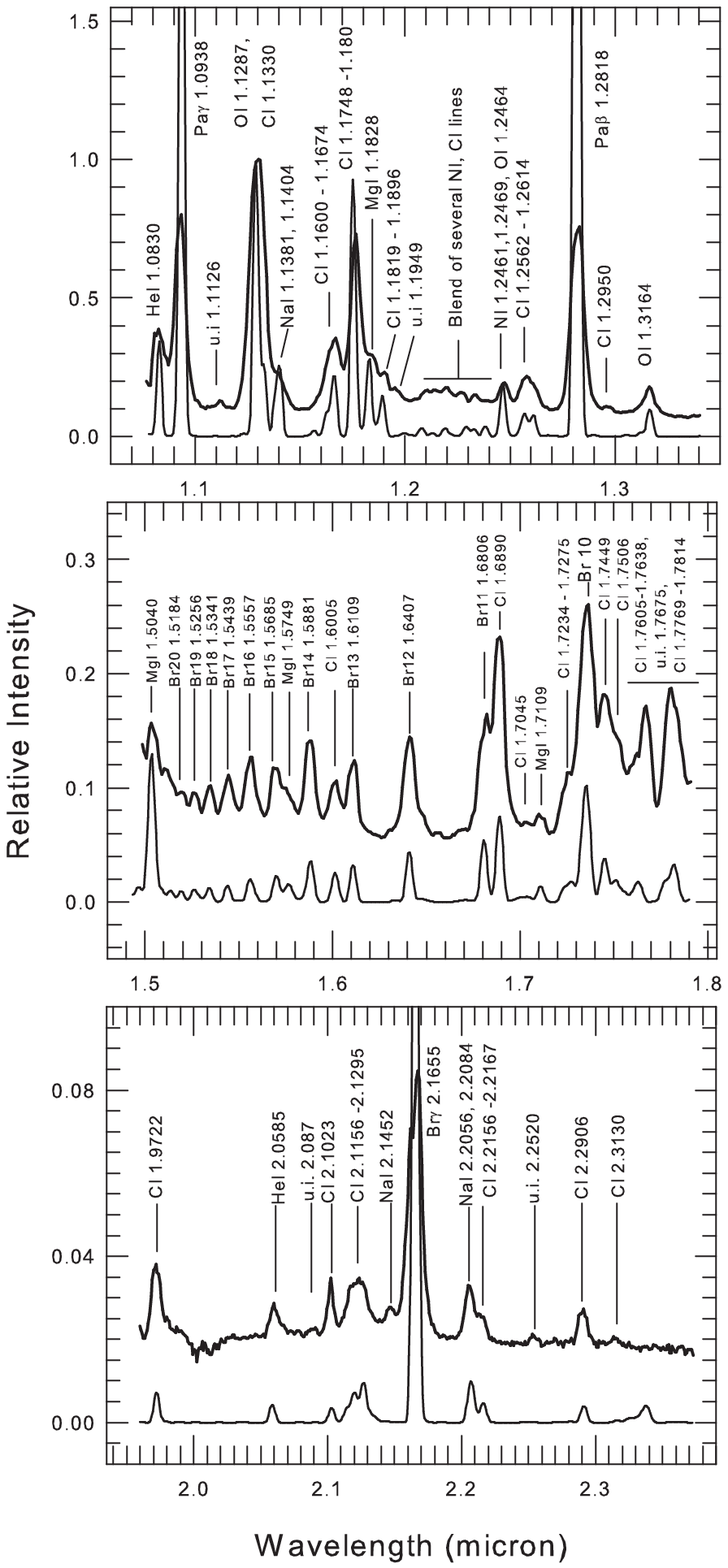}
\caption[]{ Line identification in the $J,H$ $\&$ $K$ bands shown from top to bottom
respectively.
In each panel, the
upper plot (darker shade) is the observed data of 1 March; the lower curve (lighter shade)  is the synthetic spectrum (details in section 3.3). More details on the lines are
given in Tables 3 $\&$ 4. The observed spectrum was calibrated with the observed
$JHK$ magnitudes and normalized to unity w.r.t the 1.1287 ${\rm{\mu}}$m O I line
whose peak strength is 5.6$\times$10${^{\rm -15}}$ W/cm${^{\rm 2}}$/${\rm{\mu}}$m. The absorption features at $\sim$ 2.0 ${\rm{\mu}}$m are residuals from improper telluric
subtraction (discussed in Section 2).
}
\label{fig2}
\end{figure}

6) The  C I feature at $\sim$ 2.12 ${\rm{\mu}}$m  is actually a blend of several C I
lines; the principal ones being  at 2.1156, 2.1191, 2.1211, 2.1260 and
2.1295 ${\rm{\mu}}$m.The superposition of these closely spaced lines gives the
overall feature its unusually broad appearance.

7) There are a few weak lines which remain  unidentified. One of these
is the 1.1126 ${\rm{\mu}}$m line
which appears consistently in our spectra and which is often seen in the spectra of
novae; Rudy
et al. (2003) suggest its association with Fe II. However,our model  shows that if this line is due to
 Fe II then several other lines of Fe II should also be seen - in the $H$ band especially.
Hence we are doubtful about its origin. Similarly, a weak line is consistently seen at
2.2520 ${\rm{\mu}}$m, prominently present too in the spetrum of V705 Cas (Evans et al. 1996),  which eludes identification but which may be due to  CO${^{\rm +}}$ (Dalgarno et al. 1997)

It may be noted that the Hydrogen line strengths are considerably overestimated in our model,
 particularly the stronger lines viz. Pa $\beta$ and   Br $\gamma$.
One possibility, that could cause such behavior, is that the assumption that  LTE conditions
prevail is a simplification. But this could be resolved if optical depth effects are
considered. To illustrate this, we note that even if  a non-LTE situation is considered,
as in Case B computations, then  some difficulty may still persist in explaining the
 observed relative strengths of the H lines. For e.g., Pa $\beta$ at
1.2818 ${\rm{\mu}}$m is seen to be weaker than Pa $\gamma$ at 1.0938 although Case B predicts the other way
around (typically  Pa$\beta$/Pa$\gamma$ = 1.6 is expected). As we understand it, the optical depth in a line is proportional to the column
density and the oscillator strength of the line. While the column density for all H
lines will be the same, the oscillator strength decreases considerably for the weaker
lines in a series (i.e. it   is lower for Pa $\gamma$ as compared to Pa $\beta$). This
effectively  causes the strength of   Pa $\beta$ to be reduced much more than for
Pa $\gamma$. Rigorous modeling demonstrating this effect is given in Lynch et al. (2000) and
an observed example is seen in nova N Sgr 2001 (Ashok et al. 2006) where Br $\gamma$ is seen to
be much weaker than later Br series lines like Br 10, 11 etc.
In the present case optical depth effects seem to be present, which if taken into account,
is quite likely to reduce the strength of the stronger H lines and bring them in
better agreement with the observed spectrum. In comparison to H, the optical depth
in the lines of the other elements is expected to be significantly less - and hence their strengths
remain unaffected -  because of their considerably lower column densities (as a
consequence of their lower abundances).

It was also found that use of a single temperature for the emitting gas  fails to
simultaneously reproduce the observed strengths of lines from all the elements.
Though a complex interplay between the Saha and Boltzmann equations is involved,
one of the principal reasons for this failure we believe,  can be traced to the
considerable diversity
in ionization potentials (I.P) of the elements contributing to the spectrum. The
observed lines
in the spectrum  are from neutral species. If a higher temperature
is considered ($\sim$ 10,000K), then the ionization equation indicates that elements
like Na and Mg with low I.Ps of 5.139 and 7.646 eV respectively have a larger fraction
of their atoms in higher stages of ionization and very few in the neutral state. Therefore the
lines from the neutral species of these elements are  found
to be extremely weak - the reduction in strength is
compounded by the additional fact that they have low abundances. In comparison
to Na and Mg,  higher temperatures favor
a relatively larger fraction of neutral species for H, C, N and O  because of
their  significantly higher I.Ps (13.6, 11.26, 14.53 and 13.62 eV
respectively). On the other hand, instead of a high temperature, the use of only a
single lower temperature ($\sim$ 3500-4500K) enhances
the strength  of Na and Mg lines to an unacceptable extent - they  become too strong
vis-a-vis lines of other elements. \\

It therefore becomes necessary to consider the possibility of
temperature variation in the ejecta. We have adopted the simplest
 scenario  viz. there exists  a hot zone in the ejecta outside which lies a relatively
 cooler
 region. This involves  invoking the least number of free parameters i.e. two temperature
 values for the hot and cool zone. Physically, the assumption of  temperature variation
 in the nova shell does not appear  unduly unreasonable - the formation of dust in V1280
 Sco does indicate that there  must a  cool  region in the ejecta - though the temperature
 stratification is expected to be more complex than being characterized by just two
 temperature values. But our aim, as mentioned earlier, is to get a better qualitative
 idea of the characteristics of the emission lines seen. The model computation in
 Figure 5 is therefore  made on this basis using temperatures of 8300 and 3800K  for the
 hot and cool zones with their respective emitting volumes being in the ratio of
 $\sim$ 3:1.   All other parameters are
 assumed to be the same in both zone viz. abundances, electron density etc. In this manner we are able to reproduce
 reasonable agreement between observed and computed lines strengths of
 all elements except Hydrogen (which as explained earlier could be significantly affected by opticla depth effects).
 The abundances that we used in Figure 5, in terms of mass fractions, are 0.41, 0.21,
0.147, 0.096, 0.13, 0.0051 and 0.0010 for H, He, C, N, O, Ne and Na-Fe respectively.
 which
 are reasonably consistent  with that expected in CO novae (e.g. model CO4 of Jose
$\&$ Hernanz, 1998). However, we do not intend to stress
 that our model calculations  determines elemental
 abundances  in V1280 Sco or validates the assumption of LTE prevailing. But
 it appears  that departure from LTE conditions may not be too severe in the early stages  when the density in the ejecta is high.
 A conclusion  that emerges from the analysis, and of which we
 feel fairly  convinced,  is the following. The presence  of  lines of particularly
Na  and also Mg,  associated as they are with low excitation and ionization
conditions,   necessarily implies the  existence  of a cool zone. Such a zone  is conducive
for dust formation.  In the coming subsection, using observational evidence for corraboration, we   validate  the claim that whenever these lines are seen it is also  likely to be accompanied by dust formation in the nova.

\subsection{Can we predict which novae will form  Dust from early infrared spectra?}
Considerable attention has been paid to understanding the dust formation process in
novae and the fundamental physical conditions necessary for dust grain formation
(Gehrz 1988, Evans 2001). It would appear that several parameters are involved in
determining whether a nova has the ability to produce dust. Correlations between
dust formation and parameters like ejection  velocity and speed class have been
examined but it is still not completely understood which novae will  go on to form
dust. Gehrz (1988) shows that for dust formation to take  place efficiently,   it is
necessary that a sufficiently high  particle density  be available at the dust
condensation point - the high density being necessary to  enable the nucleation of
grains to take place. It is also evident that low temperatures must prevail
in the dust forming zone to enable dust to condense.  Dust may fail to form in novae
with low metal abundances. PW Vul and HR Del are examples
of dust-poor novae which probably had low CNO abundances - probably solar-like - whereas
the  novae that have formed thick dust shells have had an enrichment of  CNO elements
(Gehrz 1988 and references therein). Rudy et al. (2003) have found a significant
correlation between  novae in which CO (carbon monoxide ) emission is seen
and their dust producing capacity. They consider the known novae in which the first
overtone of CO has been detected viz. V2274 Cyg, NQ Vul, V842 Cen, V705 Cas, V1419
Aql (Rudy et al. 2003, Ferland et al. 1979, Hyland $\&$ Mcgregor 1989, Wichmann et
al. 1990; Evans et al. 1996; Lynch et al. 1995). All of these novae produced prolific
amounts of dust. Rudy et al. (2003) comment that the association of dust with CO
may not be totally unexpected, given that CO formation maybe a precursor for dust
formation (Rawlings 1988; Evans 1996) since CO emission could cool the regions
sufficiently that dust formation can occur. Additionally, the low-temperature,high
density, metal-rich environment needed for CO production also favors the production
of dust.\\

Among these physical parameters,  the signature
of a low-temperature condition  can also be inferred from
the presence of Na and
Mg lines. We have therefore inspected the  $JHK$ spectra of these  novae
to see whether Na and Mg lines were present in their spectrum. In addition, since
high densities in the ejecta seems a pre-requisite for dust formation, only the
early spectra were inspected because  the density is expected to be high at this stage before expansion thins it. Beginning with  V2274 Cyg, this object showed
lines of Na I 1.1381, 1.1404, 2.2056, 2.2084 ${\rm{\mu}}$m and the Mg I 1.5040 and 1.7109 ${\rm{\mu}}$m lines are also
detected (Rudy et al. 2003) in spectra taken 18 days after outburst. The 1.1828 and
1.5749 ${\rm{\mu}}$m Mg I lines are not seen but that appears to be due to inadequate
resolution in distinguishing them from nearby lines with which they could be blended (see Tables 3, 4 ). In V1419 Aql,
during the early stages, among the $JHK$ spectra only the early $J$ band spectrum is available taken 18 days
after maximum light - the  Mg I 1.1828 line is clearly detected in it (Lynch. et al. 1995).
1.6-2.2 ${\rm{\mu}}$m  spectra of NQ Vul were recorded  by Ferland et al (1979) 19 days after
the outburst. The S/N in these spectra is low but the Na I 2.2056, 2.2084 ${\rm{\mu}}$m features are
clearly discerned and listed as observed features. V705 Cas was extensively studied by Evans et al (1996) with $K$ and $L$ band spectra
recorded at three epochs prior to strong dust formation in this nova which took place
$\sim$ 62 days after
outburst.  V705 Cas is an archetypal example in which dust formation was accompanied by a drastic drop of $\sim$ 7 magnitudes in the visual light curve. The $K$ band spectra of V705 Cas taken a day before maximum and 26.5 days after peak light
very prominently show the lines of Na I   2.2056,  2.2084 ${\rm{\mu}}$m  respectively and also the
line at 2.1045 ${\rm{\mu}}$m  which is attributed to Na I (Evans et al. 1996).
In the case of V842 Cen, while 1-5 ${\rm{\mu}}$m  IR spectra have certainly been recorded (Hyland $\&$ Mcgregor, 1989; Wichmann et al. 1990,1991, we are unable to examine the  $JHK$ region of the spectra - to infer about the presence of specific Na I/Mg I lines - as they do not
appear to be published. The published Hyland $\&$ Mcgregor (1989) spectra cover  the 2.9-4.1 ${\rm{\mu}}$m  region emphasizing the PAH features detected in this while the  Wichmann et al (1990) spectrum covers
only  the 2.27-2.43 ${\rm{\mu}}$m  region i.e. the focus is on the  CO first overtone
emission seen in the object. However Wichmann et al. (1991) do report that emission lines of Na I were seen, apart from other lines of H, He, C, O, N, in the 1-4 ${\rm{\mu}}$m region. Thus all these four dust forming novae had lines of Na I and/or Mg I in their spectra. \\

Apart from these novae, we were unable to locate
early $JHK$ spectra of other novae to increase the sample size to test our hypothesis.
However, we have some unpublished data on V2615 Oph, V476 Scuti and V2575 Oph
(Das et al. 2008; preliminary results in Das et al. 2007a, b, c) - it may be noted that strong
 first overtone CO emission was observed  in V2615 Oph (Das et al. 2007c).
Preliminary results indicate all three novae formed dust - from sharp declines seen in their light-curves accompanied by the
buildup of an infrared excess. Quite consistently, Na I and Mg I lines are seen in the spectra of these objects too. And finally, they are seen in  the present case of V1280 Sco; though it must be noted that CO did not form in this nova and yet it formed dust. Thus there appears to be a extremely good one-to-one correlation between the presence of NaI/MgI lines and the dust forming potential of a nova. Their presence, in early infrared spectra, could
therefore predict dust-formation. All of these novae appear to be rich in heavy
elements. Abundance calculations show  this for V842 Cen (Andrea et al. 1994)
while for the other novae, for which abundance calculations are  unavailable,  a metal-rich ejecta can be inferred indirectly from the prominent/strong  lines
of C, N, O seen in their $JHK$ spectra (especially the lines of Carbon).
In essence, dust formation in novae can be expected when conditions of
low temperatures and high densities are satisfied along with the possible
pre-requisite of an enhancement in heavy elements.

\begin{table}
\caption[]{List of observed lines in the $JHK$ spectra}
\begin{tabular}{llrr}
\hline\\
Wavelength         	& Species             	& Other contributing& \\
(${\rm{\mu}}$m)    	&                     	& lines $\&$ remarks& \\
\hline
\hline \\
1.0830             	& He \,{\sc i}        	& 				&\\	
1.0938   	   	& Pa $\gamma$     	&        			& \\
1.1126   		& u.i               & Fe \,{\sc ii}? 				&\\
1.1287   		& O \,{\sc i}      	& 				&\\
1.1330   		& C \,{\sc i}      	& 				&\\
1.1381   		& Na \,{\sc i}     	& C I 1.1373			&\\
1.1404   		& Na \,{\sc i}     	& C I 1.1415			&\\
1.1600-1.1674 		& C \,{\sc i}      	& strongest lines at		&\\
                        &                 	&1.1653, 1.1659,1.16696		&\\
1.1748-1.1800 		& C \,{\sc i}      	& strongest lines at		&\\
                        &                 	&1.1748, 1.1753, 1.1755		&\\
1.1828			& Mg \,{\sc i}           &                              &\\
1.1819-1.1896  		& C \,{\sc i}         	&strongest lines at             &\\
			&			    &1.1880, 1.1896          	&\\
1.1949              & u.i           &             &\\
1.2074,1.2095   	& N \,{\sc i}    &blended with C I 1.2088     & \\
1.2140   		& u.i            &  						&\\
1.2187,1.2204  		& N \,{\sc i}    &                          &\\
1.2249,1.2264  		& C \,{\sc i}    & 						&\\
1.2329   		& N \, {\sc i}   &  &\\
1.2382   		& N \,{\sc i}    &   &\\
1.2461,1.2469  		& N \,{\sc i}    &blended with O I 1.2464 &\\
1.2562,1.2569  		& C \, {\sc i}   &blended with O I 1.2570    &\\
1.2601,1.2614   	& C \, {\sc i}   &     &\\
1.2659   		& u.i            &  &\\
1.2818   		& Pa $\beta$         	& 				&\\
1.2950   		& C \,{\sc i}        	&				&\\
1.3164   		& O \,{\sc i}		&                               &\\
1.5040              & Mg \,{\sc i}          & blended with Mg I &\\
                    &                       &1.5025,1.5048    &\\
1.5184   		& Br 20                 &                 		&\\
1.5256   		& Br 19                 &   				&\\
1.5341   		& Br 18              	&  				&\\
1.5439   	   	& Br 17              	&  				&\\
1.5557   		& Br 16              	&   				&\\
1.5685  		& Br 15             	&    				&\\
1.5749              & Mg \,{\sc i}          &blended with  Mg I &\\
                    &                       &1.5741,1.5766, C I 1.5784 &\\
1.5881   		& Br 14              	& blended with C I 1.5853				&\\
1.6005  		& C \,{\sc i}           & 				&\\
1.6109    		& Br 13    		&  				&\\
1.6335   		& C \,{\sc i}   	&  				&\\
1.6419  		& C \,{\sc i} 		&  				&\\
1.6407  		& Br 12       		&blended with  C I lines &\\
                    &                   &between 1.6335-1.6505&\\
1.6806   		& Br 11    		&  				&\\
1.6890   		& C \,{\sc i}   	&  				&\\
1.7045   		& C \,{\sc i}   	&  				&\\
1.7109                  & Mg \,{\sc i}    	&  				&\\
1.7200-1.7900  		& C \,{\sc i}       	&Several C I lines in 		&\\
			&			&this region (see Fig.5) 			&\\
1.7362  		& Br 10      		&Affected by C I 1.7339 		&\\
			&			&line 				&\\
1.9722   		& C \,{\sc i}    	&  				&\\
2.0585 			& He \,{\sc i}          &  				&\\
2.0870			& u.i                   &  				&\\
2.1023 			& C \,{\sc i} 		& 				&\\
2.1138  		& O \,{\sc i}    	&This line may be present		&\\
\hline
\end{tabular}
\end{table}

\begin{table}
\caption[]{List of lines continued from Table 3}
\begin{tabular}{llrr}
\hline\\
Wavelength         	& Species             	& Other contribut-&\\
(${\rm{\mu}}$m)    	&                     	& ing lines/remarks		&\\
\hline
\hline \\
2.1156-2.1295  		& C \,{\sc i}     	&blend of several C I 		&\\
                        &                       &lines strongest being 		&\\
			&			& 2.1156,2.1191,2.1211, 	&\\
			&			& 2.1260,2.1295  		&\\
2.1452   		& Na \,{\sc i}?   	&  				&\\
2.1655   		& Br $\gamma$     	&  				&\\
2.2056   		& Na \,{\sc i}          & 				&\\
2.2084  		& Na \,{\sc i}          & 				&\\
2.2156-2.2167		&            		&blend of C I lines at 		&\\
			&			&2.2156,2.2160,2.2167  		&\\
2.2520   		& u.i 			&				&\\
2.2906   		& C \,{\sc i}      	& 				&\\
2.3130   		& C \,{\sc i} 		& 				&\\
\hline
\end{tabular}
\end{table}

\subsection{Evidence for sustained mass loss}
Examination of the Br$\gamma$ 2.1655 ${\rm{\mu}}$m  feature in the data of 24, 25, 27 Feb and
01 March (Figure 4) shows that there appears to be an additional line
on the blue wing of Br$\gamma$. However, we are not able to  identify this feature
with any known line and therefore looked for alternative reasons for its origin. As it turns out, it is caused by the continued presence
of the absorption component of a P-Cygni feature that persists quite
long after the outburst began - atleast till 01 March. To demonstrate this,
a velocity plot of Br$\gamma$ region is shown in Figure 6 on different days.
The earliest spectra of Feb. show classic  P-Cygni profiles. What may be noted
however is that the position of the P-Cygni absorption component in the earliest spectra closely tracks
the trough seen in the peak of the Br gamma line at later stages (Feb 24 to
March 01). The trough it may be seen, by splitting the peak/profile of the Br gamma line into
two components,  creates the impression of an additional line
being present which however is  not the case. The profiles in Figure 6 can instead be interpreted as follows. A P-Cygni profile arises in an outward
accelerating wind and is a indicator of mass loss. The initial P-Cygni structure arises from mass loss during the rise to maximum or epochs close to it. As this matter moves away from the central remnant, the line profile associated with it is expected to become  purely of  emission-type. However if strong mass loss
continues, then superposed on this emission component will be the P-Cygni component arising from the continuing mass loss. The addition of these two components, as can be  visualized, can lead to the profiles of the observed shape in Figure 6. In effect, Figure 6 indicates that observational evidence is seen that  mass-loss continues from the
nova up to 1-3 March 2007 i.e. it lasts for atleast  25-27 days after the outburst began. This estimate for the duration of the mass-loss ($T$${_{\rm ml}}$) is likely to be a lower limit for the process of mass ejection  could continue longer but at reduced mass-loss rates insufficient to create the necessary absorption column density to produce a discernible P-Cygni absorption component on the overall profile. Our finding gives direct observational evidence for sustained mass-loss and a
constraint on its duration. \\

A comparison could  be made  between theoretical estimates of $T$${_{\rm ml}}$ and its observed value from the grid of nova models developed by Prialnik and Kovetz (1995) and Yaron et al. (2005) providing numerical estimates
for key quantities characterising a nova outburst and properties of the ejecta. The different models are computed by varying the three basic and independent parameters that control a nova outburst viz the mass ($M$${_{\rm WD}}$) and temperature of the WD and the  mass accretion rate on the WD. One of the central timescale parameters that they compute   is $T$${_{\rm ml}}$ which is shown to be of the order of $t$${_{\rm 3}}$ (the time for a decline of 3 magnitudes in the visual).
The early formation of dust in V1280 Sco makes it difficult to estimate $t$${_{\rm 3}}$. But
if we proceed along the lines as our estimate of $t$${_{\rm 2}}$, then $t$${_{\rm 3}}$ should be in the range of
30-40 days. This seems to be  satisfactorily consistent with the observed lower limit for the mass-loss timescale of about 25-27 days. Together with the observed estimate of $T$${_{\rm ml}}$  if the known outburst amplitude ($A$) and the expansion velocity ($V$${_{\rm exp}}$ ) are also considered; then the grid of models can be used to estimate $M$${_{\rm WD}}$  since $A$ and $V$${_{\rm exp}}$ are two other observable parameters predicted by the computations. In this context, it appears that a value of $M$${_{\rm WD}}$
towards the higher end (1-1.25 $M$${_{\odot}}$) may be supported. For e.g. one of the models
with  $M$${_{\rm WD}}$ = 1.25 $M$${_{\odot}}$  predicts a  maximum velocity,  average velocity, outburst amplitude and $T$${_{\rm ml}}$ with values of  1230 km/s,
678 km/s, 14.3 magnitudes and 22 days respectively. This agrees reasonably with the $A$
and $V$${_{\rm exp}}$ values observed in V1280 Sco. However, this estimate of $M$${_{\rm WD}}$ has to viewed with due caution given the uncertainties in the observable quantities (we have only a lower limit on $T$${_{\rm ml}}$ and on $A$) in addition to the complicated behaviour of V1280 Sco which shows evidence for a  second episode of significant mass loss about $\sim$ 100 days after the first outburst (see  Section 3.7). It is also noted that a high WD mass would
imply a ONeMg rather than a CO WD in this nova.  But the infrared spectrum definitely supports a CO classification. Thus there could be a possible inconsistency here, which though not firmly established,  could still  be kept in mind.

\begin{figure}
\centering
\includegraphics[bb=0 0 289 297,width=3.2in,height=3.2in,clip]{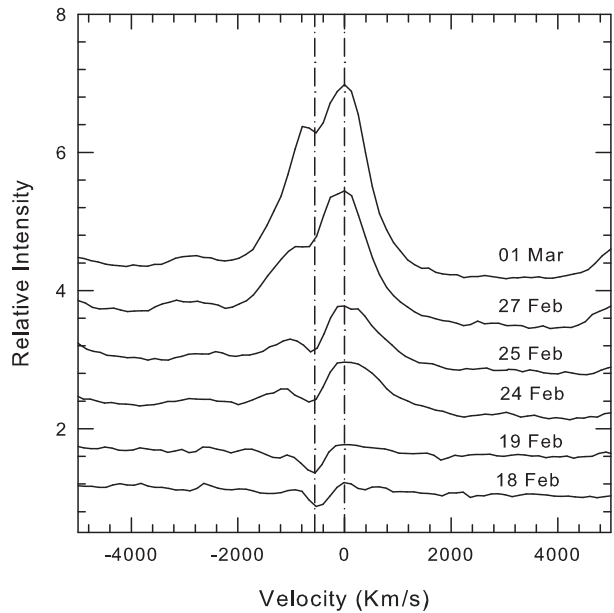}
\caption[]{ The time evolution of the Brackett $\gamma$ 2.1655${\rm{\mu}}$m line
showing the P-Cygni profile in the early stages with its absorption feature
at $\sim$ -575 km/s blueward of the emission peak.  An absorption component, coincident
in wavelength with the P-Cygni absorption trough position, persists near the peak of the profile for several days indicating continuing
mass loss (see text of section 3.5 for details).}
\label{fig2}
\end{figure}


\begin{figure}
\centering
\includegraphics[bb=76 404 360 778,width=3.2in,height=4.5in,clip]{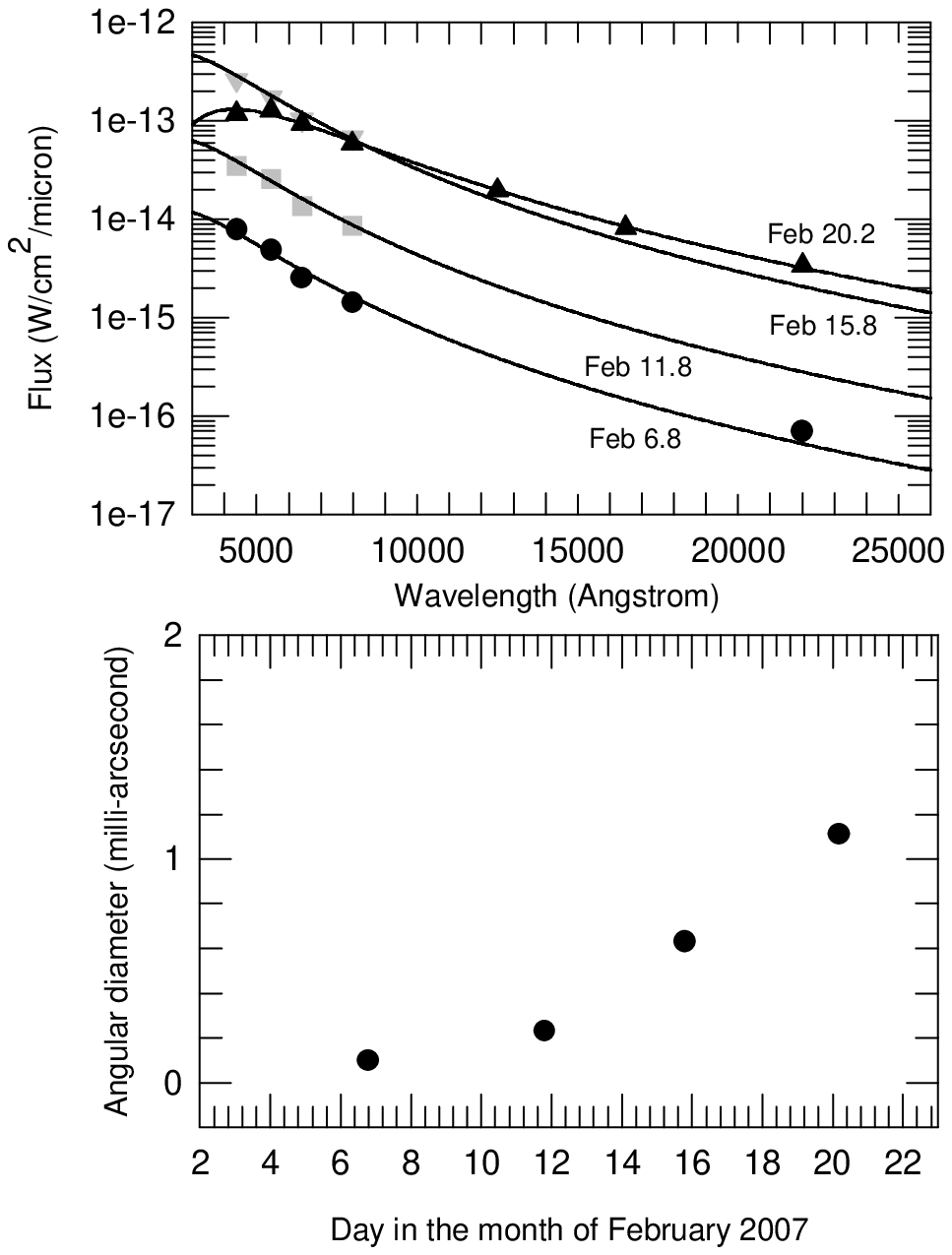}
\caption[]{ The top panel shows the observed SED's on different dates fitted by
black body fits after correcting for $A$${_{\rm v}}$ = 1.2. The same black
body temperature
of 11000K is found to fit the data of Feb 5.8, 11.8, and 15.8 while for
Feb. 20.2 a temperature of 6750K is used. The bottom panel shows the fireball
expansion in V1280 Sco during pre-maximum and peak phase.}
\label{fig1}
\end{figure}


\begin{figure}
\centering
\includegraphics[bb=0 0 416 409, width=3.2in,height=3.2in,clip]{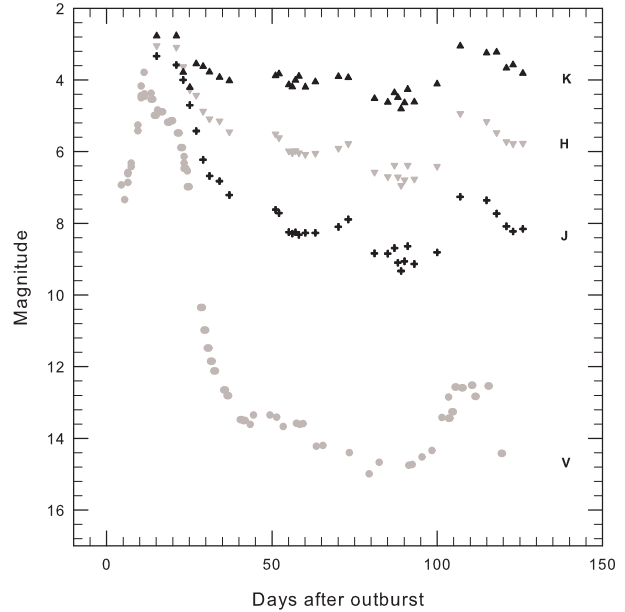}
\caption[]{ The $JHK$ lightcurve of V1280 Sco from Mt. Abu observations overlaid on
the $V$ band lightcurve for comparison ($V$ band - grey circles; $J$ band - black plus signs; $H$ band - grey downward triangles; $K$ band - black upward triangles).}
\label{fig2}
\end{figure}


\subsection{The initial fireball phase of V1280 Sco}

V1280 Sco was first caught brightening on Feb 4.5 (Sakurai $\&$ Nakamura 2007) when it was found to be at 9.4-9.9 magnitudes in the visible. Within the next $\sim$ 12 days it brightened by
nearly 5.8 mags to reach its maximum on Feb 16.2 at V=3.79 (Munari et al. 2007). During this prolonged rise to maximum, extensive optical photometry of the object has been documented;  not many novae have had been studied so well in the pre-maximum rise.  Thus the available data  permits a study to be made of the early fireball expansion in the nova along similar lines as in PW Vul (Gehrz et al. 1988). In Figure 7,
using $B$,$V$,$R$${_{\rm c}}$,$I$${_{\rm c}}$ and  $JHK$ magnitudes, when available, reasonably good blackbody fits are  obtained to the data thereby allowing a  blackbody temperature $T$${_{\rm bb}}$ to be derived. The blackbody temperatures obtained are consistent with the  observation that the nova pseudosphere generally shows an A to F spectral type at outburst.The blackbody angular diameter
$\theta$${_{\rm bb}}$ in arcseconds is then obtained from (e.g. Ney $\&$ Hatfield, 1978)

\begin{equation}
\theta{_{\rm bb}}= 2.0\times{10{^{\rm 11}}}{{({\lambda}{F{_{\lambda}}})}{_{\rm max}}}{^{\rm 1/2}}{T{_{\rm bb}}}^{\rm -2}
\end{equation}

where ($\lambda$F${_{\lambda}}$)${_{\rm max}}$ is in W/cm${^{\rm 2}}$ and $T$${_{\rm bb}}$ in Kelvin.
The estimate of  $\theta$${_{\rm bb}}$ will always be a lower limit since it is
applicable for a black body (Ney $\&$ Hatfield, 1978; Gehrz et al. 1980). For a gray body, the actual angular size can be larger,
since the right hand side of  equation 1 should then be  divided by $\epsilon$${^{\rm 1/2}}$ , where $\epsilon$ the emissivity has a value less than unity for a gray body. \\

The angular diameter
values are plotted in the lower panel of Figure 7 and clearly show the expansion of
the fireball. Though subject to some uncertainty, it would appear the expansion
rate was marginally slower between 6 to 12 Feb. but subsequently increased therafter
to $\sim$ 0.105 mas/day. This angular expansion rate is  approximately a factor of three smaller than the mean rate of 0.35 mas/day determined from the interferometric data.
As a consequence of this, values of parameters such as the  angular size and the  distance to the nova,
when derived from the interferometric and fireball approach will not match.
 For e.g the March 23 interferometric
data for which  the shell size was well constrained, estimates its diameter to be 12-13 mas in comparison to $\sim$ 4.1 mas obtained by extrapolating the data in  Figure 7. Similarly, the Chesneau et al. (2008) results estimate a mean
distance of 1.6 kpc for an adopted expansion velocity for the ejecta of 500km/s ; a similar choice of the
expansion velocity will yield a distance three times larger using the fireball approach (since the distance is inversely proportional to the angular expansion rate).
It is thus necessary to understand why this inconsistency arises between the two methods.
The most likely source for this disagreement is that diameters derived from equation 1
are  lower limits being applicable to a true blackbody which may not be the case here. The physical reasons
which causes a departure in the nature of the emission from black to gray - as seems to be applicable   here - are not entirely clear to us. Is it associated with
clumpiness in the matter or the  matter becoming optically thin? The
fireball expansion data, extrapolated back to zero angular size, shows that the onset of the eruption began around $\sim$ 2.5  Feb, almost 2.35 days before the day of discovery
and nearly 13.7 days before maximum light. A comparison could be made with V1500 Cyg and PW Vul, two other rare instances where the early fireball was documented sufficiently  to enable estimating the onset of ejection. In both these novae, ejection began well before maximum light viz. 1.8 and 9 days for V1500 Cyg and PW Vul respectively (Gehrz 1988 and references therein).

\subsection{The $JHK$ light curve of V1280 Sco}
The $JHK$ light curve is presented in Figure 8 along with the visible light curve
to enable a comparison. The first onset of dust around 12 days after maximum shows
a sudden drop in the $V$ magnitude. The contribution from the dust to the infrared affects the near-IR
bands whose magnitudes do not drop so sharply. The contribution to the $K$ band is the greatest indicating the dust emission peaks at even longer wavelengths. That this is indeed so  can be seen from the  0.5 - 13 ${\rm{\mu}}$m SED plots of the object by Chesneau et al. (2008). These plots, though obtained at different epochs, show the SED  to generally peak at 3 ${\rm{\mu}}$m or slightly beyond indicating a blackbody temperature for the dust of $\sim$ 1000K. Such peak  temperatures for the dust are typically found in novae (Gehrz et al. 1988). Figure 8 shows that on the whole,
the $K$ band brighness remains quite high througout our observations indicating the
constant presence of dust during this time. The dust shell formed in V1280 Sco is not
as optically thick as in  novae like Nova Serpentis 1978 (Gehrz et al. 1980)
or NQ Vul (Ney $\&$ Hatfield 1978) where the bulk of the entire  emission at shorter wavelengths is
absorbed and re-radiated by the dust into the IR. By fitting blackbody curves to our
$JHK$ magnitudes and the $N$ band fluxes of Chesneau et al. (2008), we estimate for V1280 Sco, that the ratio of the maximum infrared luminosity to the outburst luminosity (from data in Figure 7)  is $\sim$ 0.06. This ratio is found to be close to unity for optically thick novae like Nq Vul, LW Ser etc. The fact that the IR luminosity is much less than the outburst
luminosity, but that the dust is thought to cause the decline in the visual band, is suggestive of a clumpy shell. The clumps are individually optically thick, but do not cover the whole of the sky as seen from the nova. The decline in the $V$ band could then occur when an optically thick clump of dust forms along the line of sight. \\

A notable feature of the light-curve behaviour is the strong brightening seen at $\sim$ 110 days after outburst. This is seen in the
$V$ band
and also $JHK$ bands. If the $V$ band brightening was to arise because of destruction
of the dust shell, then an increase in the IR fluxes would not be expected. Since this is not the case, the observed behavior possibly indicates that a second outburst or significant  episode of mass loss has taken place at this stage. Such rebrightenings,
at late stages, are occasionally known to occur in  novae (e.g. V1493 Aql;
Venturini et al. 2004) but the cause for it is open to interpretation. The interferometric results (Chesneau et al. 2008) indicate the possibility of a second dust shell forming from the matter ejected in this rebrightening. Though our IR observations could not be continued beyond 126 days after the outburst, the $V$ band photometry shows that V1280 Sco had another phase of rebrightening in October-November 2007 (Munari
$\&$ Siviero,  2007) which lasted for several days. On the whole the object has exhibited a complex light curve.

\section{ Summary} Near-infrared spectroscopy and photometry of the dust forming nova V1280 Sco are presented. The key result concerns the issue of predicting which novae will form dust. Diagnostic lines are identified  which are shown to reliably aid  in such a prediction. The robustness of the predictive power of these lines can be tested through further observations. A synthetic LTE spectrum is generated to facilitate line identification, to qualitatively study different aspects of the observed spectral features and to give estimates of elemental abundances in V1280 Sco.  The presence of  a persisting absorption structure in the Br $\gamma$ line is interpreted as evidence for sustained mass-loss during the outburst and used to set a lower limit of 25-27 days for the mass loss duration. Consistent with the nova's prolonged climb to maximum, it is shown that the actual outburst commenced very early indeed - approximately 13.7 days before  maximum light.

\section*{Acknowledgments}

The research work at Physical Research Laboratory  is funded by the Department
of Space, Government of India.  We are thankful for the online availability of
the Kurucz and NIST atomic linelist database. We  also thank the VSNET $\&$ VSOLJ, Japan;
and  AFOEV, France for the use of  their optical photometric data.

\end{document}